\documentclass[aps, nofootinbib, superscriptaddress, pra, twocolumn]{revtex4-1}
\usepackage{ORI_Group_style}
\usepackage{soul}

\newcommand{\Ob}{O\bold{e}_1\bold{e}_2\bold{e}_3}
\newcommand{\Ol}{O\uex\uey\uez}

\newcommand{\gr}{\gamma_0}

\newcommand{\Eang}{\hat{\W}}

\newcommand{\eula}{\alpha}
\newcommand{\eulb}{\beta}
\newcommand{\eulc}{\gamma}

\newcommand{\deula}{\Dot{\alpha}}
\newcommand{\deulb}{\Dot{\beta}}
\newcommand{\deulc}{\Dot{\gamma}}

\newcommand{\ppop}{\hat{\pp}}
\newcommand{\rrop}{\hat{\rr}}
\newcommand{\Sup}{\Sop_{\uparrow}}
\newcommand{\Sdw}{\Sop_{\downarrow}}

\newcommand{\Jup}{\Jop_{\uparrow}}
\newcommand{\Jdw}{\Jop_{\downarrow}}
\newcommand{\Jud}{\Jop_{\uparrow\downarrow}}
\newcommand{\D}[2]{\hat{\mathcal{D}}_{#1}^{#2}}	
\newcommand{\rop}{\hat{r}}
\newcommand{\yop}{\hat{y}}
\newcommand{\XXi}{\boldsymbol{\xi}}
\newcommand{\Xiop}{\hat{\boldsymbol{\xi}}}
\newcommand{\xiop}{\hat{\xi}}

\newcommand{\wR}{\w_I}
\newcommand{\wD}{\w_D}
\newcommand{\wL}{\w_L}
\newcommand{\wT}{\w_T}
\newcommand{\wZ}{\w_Z}
\newcommand{\wS}{\w_S}

\newcommand{\mmu}{\boldsymbol{\mu}}

\begin{document}

\title{Linear Stability Analysis of a Levitated Nanomagnet in a Static Magnetic Field: Quantum Spin Stabilized Magnetic Levitation}

\author{C. C. Rusconi}
\affiliation{Institute for Quantum Optics and Quantum Information of the
Austrian Academy of Sciences, A-6020 Innsbruck, Austria.}
\affiliation{Institute for Theoretical Physics, University of Innsbruck, A-6020 Innsbruck, Austria.}
\author{V. Pöchhacker}
\affiliation{Institute for Quantum Optics and Quantum Information of the
Austrian Academy of Sciences, A-6020 Innsbruck, Austria.}
\affiliation{Institute for Theoretical Physics, University of Innsbruck, A-6020 Innsbruck, Austria.}
\author{J. I. Cirac}
\affiliation{Max-Planck-Institut f\"ur Quantenoptik,
Hans-Kopfermann-Str. 1, D-85748, Garching, Germany.}
\author{O. Romero-Isart}
\affiliation{Institute for Quantum Optics and Quantum Information of the
Austrian Academy of Sciences, A-6020 Innsbruck, Austria.}
\affiliation{Institute for Theoretical Physics, University of Innsbruck, A-6020 Innsbruck, Austria.}

\begin{abstract}
We theoretically study the levitation of a single magnetic domain nanosphere in an external static magnetic field.
We show that apart from the stability provided by the mechanical rotation of the nanomagnet (as in the classical Levitron), the quantum spin origin of its magnetization provides two additional mechanisms to stably levitate the system. Despite of the Earnshaw theorem, such stable phases are present even in the absence of mechanical rotation.
For large magnetic fields, the Larmor precession of the quantum magnetic moment stabilizes the system in full analogy with magnetic trapping of a neutral atom.  For low magnetic fields, the magnetic anisotropy stabilizes the system via the Einstein-de Haas effect. These results are obtained with a linear stability analysis of a single magnetic domain rigid nanosphere with uniaxial anisotropy in a Ioffe-Pritchard magnetic field.

\end{abstract}

\maketitle

\section{Introduction}

According to the Earnshaw theorem~\cite{earnshaw1842nature,bassani2006earnshaw} a ferromagnet can   be    stably levitated in a static magnetic field only when it is mechanically rotating about its magnetization axis. Such a gyroscopic-based stabilization mechanism can be neatly observed with a Levitron~\cite{harrigan1983levitation,Gov1999214,Dullin,Berry_Levitron,simon1997spin}. The Earnshaw theorem does not account for the microscopic quantum origin of magnetization. For instance, a single neutral magnetic atom can be stably trapped in a static magnetic field by means of the Larmor precession of its quantum magnetic moment~\cite{Sukumar1997,Sukumar2006}. In both the Levitron and the atom, the magnetization, initially anti-aligned to the magnetic field, adiabatically follows the local direction of the magnetic field, thereby confining the center-of-mass motion~\cite{Berry_Levitron}.   

In this article, we study the stability of a levitated single magnetic domain particle (nanomagnet) in a static magnetic field. The magnetization of the nanomagnet couples to its center-of-mass motion via the interaction with the external inhomogeneous magnetic field, and to its orientation via the magnetocrystalline anisotropy \cite{chikazumi,O'Keeffe20122871}. The latter induces magnetic rigidity, namely its magnetic moment cannot freely move with respect to a given orientation of the crystal structure of the nanomagnet.  Together with the quantum spin origin of the magnetization, given by the gyromagnetic relation, this leads to the well-known Einstein-de Haas effect \cite{EinsteindeHaas}. That is, a change of magnetization is accompanied by mechanical rotation in order to conserve total angular momentum. The Einstein-de Haas effect is boosted at small scales due to the small moment-of-inertia-to-magnetic-moment ratio \cite{Chud_EdH1,Chud_EinsteindeHaas, MagnetoMech_Torsion}. 

We shall argue that the quantum spin origin of magnetization opens the possibility to magnetically levitate a non-rotating nanomagnet in a static field configuration. Indeed, we encounter two stable phases of different physical origin. The {\em atom}  (A) phase  appears at sufficiently large magnetic fields where the nanomagnet effectively behaves as a soft magnet, namely its magnetization can freely move with respect to its orientation. The {\em Einstein-de Haas} (EdH) phase appears at sufficiently small magnetic fields where the nanomagnet effectively behaves as a hard magnet, namely the magnetization sticks to the crystal. The EdH phase requires the magnet to be sufficiently small. Furthermore, we also recover the {\em Levitron} (L) phase for a larger rotating magnet, which can be predicted without accounting for the quantum spin origin of the magnetization. Such a rich stability phase diagram could be experimentally tested and opens the possibility to cool the several degrees of freedom of the nanomagnet in the stable phases to the quantum regime.

This article is structured as follows.
In \secref{sec:Model}, we model a single magnetic domain nanoparticle in a static field. Both a quantum and a classical description of the model is given.
In \secref{sec:LinStab} we derive the stability criterion as a function of the physical parameters of the system.
In \secref{sec:NM_Stab} we discuss the stability diagrams and the physical origin of the different stable phases. 
We draw our conclusions and discuss further directions in \secref{sec:Conclusions}.

\section{Single magnetic domain nanoparticle in a static magnetic field}\label{sec:Model}

We consider a single magnetic domain nanoparticle in an external static magnetic field $\BB(\rr)$.
The nanomagnet is modeled as a rigid sphere of radius $R$, mass $M$, and with a magnetic moment $\mmu$.
$\BB(\rr)$ is assumed to be approximately homogeneous within the volume of the sphere such that the interaction energy between $\mmu$ and $\BB(\rr)$ can be expressed as $V_b=-\mmu\cdot\BB(\rr)$, where $\rr$ is the center-of-mass position of the sphere (point-dipole approximation).
The exchange interaction between the magnetic dipoles inside a magnetic domain is assumed to be the strongest energy scale of the problem. 
Under this assumption, $\mu \equiv |\mmu|$ can be approximated to be a constant.
The degrees of freedom of the system are hence: (i) the center-of-mass motion (described by 6 parameters), (ii) the rotational motion (described by 6 parameters), and (iii) the magnetization dynamics (described by 2 parameters) \cite{miltat2002introduction}.

The orientation of the rigid sphere is represented by the three Euler angles $\W\equiv(\eula,\eulb,\eulc)$ in the ZYZ parametrization \cite{ORI_Cos_PRB}, which specify the mutual orientation between the frame $\Ob$, fixed in the object and centered in its center of mass, and the frame $\Ol$, fixed in the laboratory.
According to this convention the coordinate axes of the frame $\Ob$ and the coordinate axes of the frame $\Ol$ are related through $(\bold{e}_1,\bold{e}_2,\bold{e}_3)^T = R(\W)(\uex,\uey,\uez)^T$, where the transformation matrix reads $R(\W)\equiv R_z(\eulc)R_y(\eulb)R_z(\eula)$, where $R_\nn(\theta)$ is the clockwise rotation of the coordinate frame (passive rotation) of an angle $\theta$ about the direction $\nn$  (see \cite{ORI_Cos_PRB} for further details).
Hereafter Latin indexes $i,j,k,\ldots = 1,2,3$ label the body frame axes while Greek indexes $\mu,\nu,\lambda\ldots =x,y,z$ label the laboratory frame axes. 

Ferromagnetic materials exhibit magnetocrystalline anisotropy, namely they magnetize more easily in some directions than others, due to the interaction between the magnetic moment and the crystal structure of the material \cite{chikazumi}. 
This interaction determines preferred directions along which the magnetic energy of the system is minimized.
We consider uniaxial anisotropy, for which the preferred direction is a single axis (easy axis) in the crystal.
By choosing $\bold{e}_3$ to be the easy axis, the uniaxial anisotropy energy is given by $V_a \equiv -k_aV\pare{\bold{e}_3\cdot \mmu/\mu}^2$, where $k_a$ and $V$ are, respectively, the anisotropy energy density and the volume of the nanomagnet. $V_a$ has two minima corresponding to $\mmu$ being aligned or anti-aligned to $\bold{e}_3$.
Note that $\bold{e}_3$ depends on $\W$, and hence $V_a$ couples the magnetization with the orientation of the nanomagnet.

Regarding $\BB(\rr)$, we consider the Ioffe-Pritchard field given by \cite{atomchip}
\be\label{eq:Ioffe_Pritchard}
\begin{split}
	\BB(\rr) =&\,\uex \pare{B' x-\frac{B''}{2}xz}-\uey\pare{B' y+\frac{B''}{2}zy}\\
	& +\uez\spare{ B_0 + \frac{B''}{2}\pare{z^2-\frac{x^2+y^2}{2}}},
\end{split}
\ee
where $B_0,B'$ and $B''$ are the three parameters characterizing the Ioffe-Pritchard trap, namely the bias, the gradient, and the curvature. 
This field, which is commonly used to trap magnetic atoms \cite{atomchip}, is non-zero at its center, \ie~$\BB(0)=B_0\uez$. Gravity is assumed to be along the $z$-axis. This shifts the trap center from the origin $\rr=0$ along $z$ by an amount $M g/(\mu B'')$, where $g$ is the gravitational acceleration. Provided this shift is smaller than the length scale $(B_0/B'')^{1/2}$ ($B'/B''$) over which the Ioffe-Pritchard field significantly changes along $z$ on-axis (off-axis), gravity can be safely neglected. In the parameter regime considered in this article, this is always the case. Indeed, we have checked that the stability diagrams shown do not change when gravity is included. Gravity is hence neglected in the analysis hereafter. Finally, we remark that since both $M$ and $\mu$ scale with the volume, the condition to neglect gravity is the same as for a magnetically trapped atom.

In summary, our model, whose physical parameters are listed in \tabref{TAB:PhysParam}~\footnote{The physical parameters listed in \tabref{TAB:PhysParam} should not be confused with the 14 dynamical parameters describing the degrees of freedom of the nanomagnet.},  assumes a: single magnetic domain, rigid body, spherical shape, constant magnetization, uniaxial anisotropy, Ioffe-Pritchard magnetic field, and point-dipole approximation. In \secref{sec:Conclusions}, we discuss these assumptions and the potential generalization of the model. 

\begin{table}
\caption{Physical parameters of the model. Whenever required, the following values are used: $\rho_M = 10^4 \text{Kg}/\text{m}^3$, $\rho_\mu = [\hbar \gr \rho_M /(50~\text{amu})] \text{J}/(\text{Tm}^3)$(where $\gr=1.760\times10^{11}\text{rad}/(\text{s T})$ is the electronic gyromagnetic ratio and amu is atomic mass unit), $k_a = 10^4 \text{J}/\text{m}^3$, $B' = 10^4 \text{T}/\text{m}$, and $B''=10^6 \text{T}/\text{m}^2$. }\label{TAB:PhysParam}
\begin{ruledtabular}
\begin{tabular}{ll}
Parameter & Description [dimension SI]\\
\hline
$\rho_M \equiv M/V$ & mass density [$\text{Kg}\times \text{m}^{-3}$]\\
$R$ & radius [$\text{m}$]\\
$\rho_\mu \equiv \mu/V$ & magnetization [$\text{J}\times \text{T}^{-1}\times \text{m}^{-3}$]\\
$k_a$ & magnetic anisotropy constant [$\text{J}\times \text{m}^{-3}$]\\
$B_0$ & field bias [$\text{T}$]\\
$B'$ & field gradient [$\text{T}\times \text{m}^{-1}$]\\
$B''$ & field curvature [$\text{T}\times\text{m}^{-2}$]
\end{tabular}
\end{ruledtabular}
\end{table}

Given a set of values in \tabref{TAB:PhysParam}, can the nanomagnet be stably levitated? To address this question, we first need to describe the dynamics of the system.
This can be done using either quantum mechanics or classical mechanics~\footnote{As shown below, the classical description is sufficient to obtain the criterion for the stable magnetic levitation of the magnet. However, we emphasize that the stable A and EdH phases crucially depend on the quantum spin origin of the magnetization. In the classical description, this is included with the phenomenological Landau-Lifshitz-Gilbert (LLG) equations describing the magnetization dynamics~\cite{miltat2002introduction}. The quantum description does not only incorporate this key fact from first principles, but will also be useful for further research directions, see \secref{sec:Conclusions}.}.

\subsection{Quantum description}\label{sec:QuantumModel}

The degrees of freedom of the nanomagnet are described in quantum mechanics through the following quantum operators.
The center-of-mass motion by $\rrop=(\xop,\yop,\zop)$ and $\ppop = (\pop_x,\pop_y,\pop_z)$, where $[\rop_\nu,\pop_\lambda]=\im \hbar \delta_{\nu\lambda}$.
The rotational motion by $\Eang=(\hat{\eula},\hat{\eulb},\hat{\eulc})$ and $\LLop = (\Lop_x,\Lop_y,\Lop_z)$, where the Euler angle operators commute with themselves, $[\Lop_\nu,\Lop_\lambda]=\im \levi{\nu\lambda\rho}\Lop_\rho$, and the commutators $[\Eang,\LLop]$, which are more involved~\cite{ORI_Cos_PRB}, are actually not required, see below.
Regarding the magnetization dynamics, the magnetic moment is given by $\hat{\mmu} = \hbar \gr \FFop$, where $\gr$ is the gyromagnetic ratio, and $\FFop$ is the total spin  of the nanomagnet (macrospin), where $[\Fop_\nu,\Fop_\lambda]=\im \levi{\nu\lambda\rho}\Fop_\rho$.
$\FFop$ is obtained as the sum of the spin of the $N$ constituents of the nanomagnet, $\FFop = \sum_{i=1}^N \FFop_i$. In the quantum description, the constant magnetization assumption can be incorporated via the macrospin approximation: the total spin is projected into the subspace with $\FFop^2 = N f(N f+1) \equiv F(F+1)$, where $f$ is the total spin of a single constituent (assumed to be identical for simplicity).
Under the macrospin approximation the magnetization dynamics can thus be described by the two spin ladder operators $\Fop_\pm \equiv \Fop_x \pm \im \Fop_y$.
The degrees of freedom of the nanomagnet can hence be represented by the 14 quantum operators $\{\rrop,\ppop,\Eang,\LLop, \Fop_\pm\}$.

In the coordiante frame $\Ol$, the quantum mechanical Hamiltonian of the nanomagnet in terms of these operators reads \cite{ORI_Cos_PRB}
\be\label{eq:Ham_Lab}
	\Hop = \frac{\hat{\pp}^2}{2M}+ \frac{\hbar^2}{2I}\LLop^2 -\hbar\gr\FFop\cdot\BB(\hat \rr)  -\hbar^2 D \spare{\bold{e}_3(\Eang)\cdot \FFop}^2,
\ee
where $I\equiv2MR^2/5$ is the moment of inertia of a sphere, and $D\equiv k_a V/(\hbar F)^2$ parametrizes the uniaxial anisotropy strength.

As discussed in \cite{ORI_Cos_PRB}, it is more convenient to express $\Hop$ in the coordinate frame $\Ob$.
This is done via the change of variables $\hat{\bold{A}}_i(\Eang) \equiv \sum_\nu R_{i\nu}(\Eang)\hat{\bold{A}}_\nu$ for $\hat{\bold{A}}=\LLop,\FFop,\BB(\rrop)$.
The operators $R_{i\mu}(\Eang)$ can be written as a function of the 9 D-matrix tensor operators $\D{mk}{1}$, where $m,k=\pm1,0$ \cite{ORI_Cos_PRB}. 
These 9 D-matrix operators are not independent. They must satisfy the following relations \cite{morrison1987guide}
\bea
	 (-1)^{k-m} \D{mk}{j} &=& \pare{\D{-m-k}{j}}^\dag, \label{eq:Constraint_Dmtrop1}\\
	\sum_m  \D{mk}{1}\pare{\D{mk'}{1}}^\dag  &=& \delta_{kk'}\mathbb{1}, \label{eq:Constraint_Dmtrop2}\\
	\sum_k  \pare{\D{mk}{1}}^\dag \D{m'k}{1} &=& \delta_{mm'} \mathbb{1} \label{eq:Constraint_Dmtrop3}.
\eea
Using the D-matrix tensor operators, the Hamiltonian in the body frame reads~\cite{ORI_Cos_PRB} 
\be\label{eq:Ham_Body_General}
\begin{split}
	\Hop =& \frac{\hat{\pp}^2}{2M} + \frac{\hbar^2}{2 I}\pare{\JJop^2 +2\Sop_{3}\Jop_{3}+\Jup\Sdw+\Jdw\Sup}\\
	& +\hbar\gamma_0 \SSop\cdot\BB(\hat \rr,\Eang)-\hbar^2D\Sop_3^2
\end{split}
\ee
by defining $\JJop \equiv \LLop - \SSop$, $\Jop_{\uparrow\downarrow} \equiv \Jop_1 \mp \im \Jop_2$, and $\Sop_{\uparrow\downarrow} \equiv \Sop_1 \mp \im \Sop_2$, where $\SSop\equiv-\FFop$ for convenience.
The D-matrix operators fulfill the following commutations rules: $[\D{mk}{j},\D{m'k'}{j'}]=0$,  $[\Jop_3,\D{mk}{j}]=k\D{mk}{j}$, and $[\Jud,\D{mk}{j}] = \sqrt{(j\mp k)(j\pm k+1)}\D{m k\pm1}{j}$, see \citep{ORI_Cos_PRB} for further details.
The Hamiltonian $\Hop$ is invariant under a rotation about the easy axis of the nanomagnet, namely $[\Hop,\Lop_3] = [\Hop,\Jop_3-\Sop_3]=0$ \cite{ORI_Cos_PRB}.
Therefore it is convenient to define $\wS \equiv -\hbar\avg{\Lop_3}/I$, which represents the rotational frequency of the nanomagnet about the easy axis $\bold{e}_3$. Note that $\avg{\Jop_3}$ for a given $\wS$ can then be written in terms of $\avg{\Sup}$ and $\avg{\Sdw}$. Furthermore, using (\ref{eq:Constraint_Dmtrop1}-\ref{eq:Constraint_Dmtrop3}) one can express $\avg{\D{mk}{1}}\in \mathbb{C}$ $\forall m,k$ as a function of $\avg{\D{11}{1}}, \avg{\D{0-1}{1}}$ and $\avg{\D{-10}{1}}$, which are given by three real independent parameters.
Hence, we define the 13 operators
\be\label{eq:Q_dof_Body}
	\hat{\XXi} \equiv\pare{\rrop,\ppop,\Jup,\Jdw,\D{11}{1},\D{0-1}{1},\D{-10}{1},\Sup,\Sdw},
\ee
whose expectation values describe the degrees of freedom of the system in the semiclassical  approximation. With this approximation, the evolution of \eqnref{eq:Q_dof_Body} as described by $\Hop$, \eqnref{eq:Ham_Body_General}, is used in \secref{sec:LinStab} to analyze the linear stability of the system for a given value of $\wS$ and the physical parameters given in \tabref{TAB:PhysParam}.

\subsection{Classical description} \label{sec:ClassDesc}

Let us now give a classical description of the system in the Lagrangian formalism.
The center-of-mass position of the nanomagnet is described by the coordinate vector $\rr=(x,y,z)$ and its orientation by the Euler angles $\W = (\eula,\eulb,\eulc)$. The direction of the magnetic moment $\mmu/\mu$
 is described by $(\phi,\theta)$, which represent, respectively, the polar and azimuthal angles in the frame $\Ol$. 
The Lagrangian of the system reads
\be\label{eq:LL_Lagrangian}
	\mathcal{L} = T_\text{cm} + T_\text{rot}+ T_\text{mag} - V_a - V_b,
\ee
where $T_\text{cm}$, which represents the kinetic energy of the center-of-mass motion, reads
\be\label{eq:LL_Tcm}
	T_\text{cm} \equiv \frac{M}{2}\pare{\Dot{x}^2+\Dot{y}^2+\Dot{z}^2}.
\ee
The rotational kinetic energy of the rigid body in the body frame coordinate system $\Ob$, reads \cite{Goldstein}
\be\label{eq:LL_Trot}
	T_\text{rot} \equiv \frac{I}{2}\pare{\deula^2+\deulb^2 + \deulc^2 + 2\deula\deulc \cos\eulb}.
\ee
$T_\text{mag}$ accounts for the kinetic energy associated to the motion of the magnetic moment, namely \cite{miltat2002introduction}
\be\label{eq:T_mu}
	T_\text{mag} \equiv -\frac{\mu}{\gr} \Dot\phi \cos\theta.
\ee 
We remark that \eqnref{eq:T_mu} leads to the phenomenological Landau-Lifshitz-Gilbert (LLG)  describing the magnetization dynamics~\cite{miltat2002introduction,ClassicalNM}.
The quantum description given in \secref{sec:QuantumModel} has the advantage to describe this from first principles.
The classical uniaxial anisotropy interaction reads 
\be\label{eq:LL_AnisotropyInterac}
	V_a \equiv -k_a V \spare{\sin\eulb\sin\theta\cos\pare{\eula-\phi}+\cos\eulb\cos\theta}^2,
\ee
where we recall that $\bold{e}_3$ coincides with the anisotropy axis.
The magnetic dipole interaction between the external field $\BB(\rr)$ and the magnetic moment $\mmu$ reads
\be\label{eq:LL_MagneticInterac}
\begin{split}
	V_b 	\equiv &  -\mu \big[B_x(\rr)\cos\phi\sin\theta + B_y(\rr)\sin\phi\sin\theta\\
	& + B_z(\rr)\cos\theta\big].
\end{split}	
\ee
Note that $V_a$ ($V_b$) couples the magnetization $\mmu$ with the orientation $\W$ (center of mass $\rr$) of the nanomagnet.

The Lagrangian $\mathcal{L}$ is independent on $\Dot\theta$, thereby $\theta$ is not an independent dynamical variable.
In the absence of dissipation, the magnetic moment $\mmu$ undergoes a constant precession around a direction determined by \eqnref{eq:LL_AnisotropyInterac} and \eqnref{eq:LL_MagneticInterac}, and thus can be described with a single precession angle \cite{miltat2002introduction}.

Furthermore, $\mathcal{L}$ is independent on $\eulc$, and thus $L_3 \equiv \partial{}\mathcal{L}/\partial\Dot{\eulc}= I\pare{\Dot{\eulc}+\Dot{\eula}\cos\eulb}$ is a constant of motion. The quantity $L_3$ represents the rotational angular momentum of the rigid sphere about the axis $\bold{e}_3$.  Once $\wS \equiv - L_3/I$ is fixed, the state of the system can thus be described by the 13 independent parameters  $\rr,\pp,\W,\deula,\deulb, \phi, \Dot\phi$. These are, roughly speaking, the classical analogs to the quantum operators \eqnref{eq:Q_dof_Body}.

\section{Linear stability analysis}\label{sec:LinStab}

\begin{figure*}
	\includegraphics[width = 2\columnwidth]{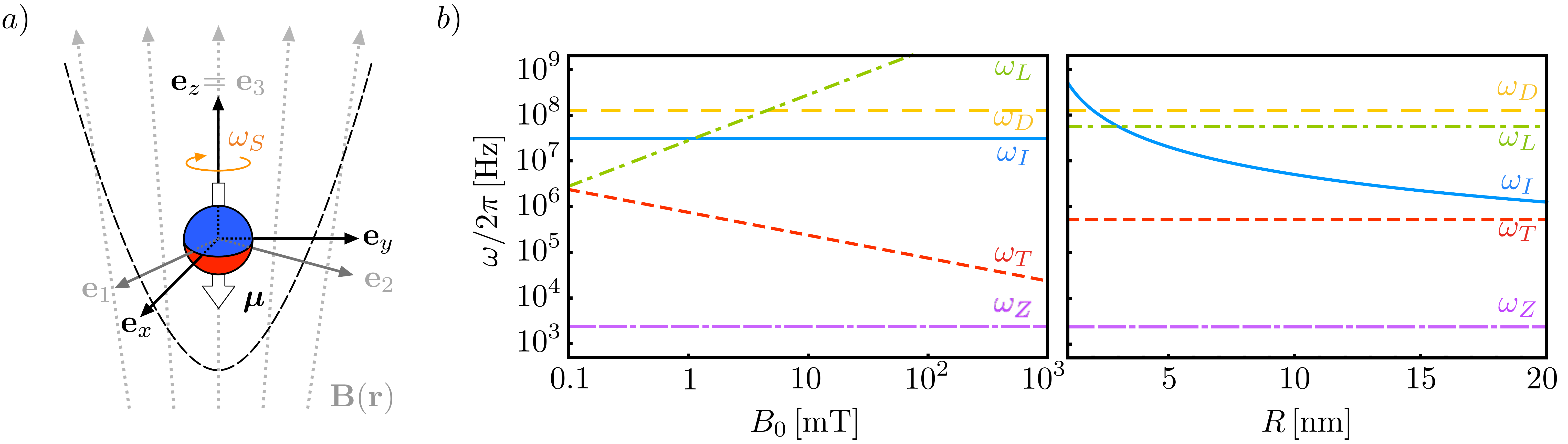}
	\caption{(a) Equilibrium solution for a levitated nanomagnet in a Ioffe-Pritchard magnetic field. The nanomagnet is at the center of the trap, rotating about $\bold{e}_3$ with angular frequency $\wS$, and the magnetic moment is parallel to the anisotropy axis and anti-aligned to $\BB(0)$. (b) $\wT,\wZ,\wL,\wR$ and $\wD$, which are defined in \tabref{TAB:Frequencies}, as a function of the bias field $B_0$ for $R=4\text{nm}$ (left panel) and of the radius $R$ for $B_0 = 2\text{mT}$ (right panel). Other physical parameters are taken from the caption of \tabref{TAB:PhysParam}. }
	\label{FIG:Frequencies}
\end{figure*}

Let us now describe the criterion which determines the linear stability of the system.
While both the classical and the quantum description lead to the same results, as discussed at the end of the section, we derive the criterion using the quantum description. 
The Heisenberg equation for the operators $\Xiop$ \eqnref{eq:Q_dof_Body} using the Hamiltonian \eqnref{eq:Ham_Body_General} can be written as $\pa{t}\Xiop=[\Xiop,\Hop]/\im\hbar \equiv \bold{G}(\Xiop)$.  $\GG$ is a vector function of $\Xiop$ that depends on the physical parameters given in \tabref{TAB:PhysParam}.
These Heisenberg equations are a non-linear system of differential equations for the operators of the system.
The stability of the system is studied in the semiclassical approximation, namely the system is considered to be in a quantum state $\hat{\rho}$ such that 
\be\label{eq:NO_Correl}
	\text{Tr}\spare{\xiop_i\xiop_j\hat{\rho}}= \avg{\xiop_i\xiop_j} \simeq \avg{\xiop_i}\avg{\xiop_j} \quad	\forall i,j,
\ee
where $\xiop_i$ is the $i$-th component of $\Xiop$. Furthermore since $\Lop_{3}$ is a constant of motion, we consider $\hat \rho$ to lie in the Hilbert subspace of eigenstates of $\Lop_{3}$ with eigenvalue $- I\w_S/\hbar$. Within this subspace one can thus use $\Lop_3 = - I \w_S \id/\hbar$. 
The Heisenberg equations of motion for a given $\w_S$ can be approximated by the closed set of semiclassical equations
\be\label{eq:Semiclass_Eq}
	\pa{t}\avg{\Xiop} = \GG( \avg{\Xiop})
\ee
by using \eqnref{eq:NO_Correl} to approximate $\avg{\GG(\Xiop)}\simeq \GG(\avg{\Xiop})$.
A solution of \eqnref{eq:Semiclass_Eq} is given by
\be\label{eq:Equilibrium_Nanomagnet}
	\XXi_0(t)\equiv\pare{\boldsymbol{0},\boldsymbol{0},0,0,\avg{\D{11}{1}}_0=e^{\im\pare{\varphi - \wS t}},0,0,0,0},
\ee
where $\varphi$ is a phase factor fixed by the initial condition on $\avg{\D{11}{1}}_0$.
This solution\footnote{Throughout this article we focus on the stability of the equilibrium solution \eqnref{eq:Equilibrium_Nanomagnet}. However, we remark that we have not exhaustively investigated the existence of other equilibrium solutions.} corresponds to a nanomagnet rotating at the frequency $\wS$ along $\bold{e}_3$, at rest in the center of the field ($\BB(0)=B_0\uez$), and with $\mmu/\mu = - \bold{e}_3 = -\uez$, namely magnetization parallel to the easy axis and anti-aligned to the field at the center, see \figref{FIG:Frequencies}.a. 

The linear stability of this solution is analyzed through the dynamics of the fluctuations $\delta\XXi(t) \equiv \avg{\Xiop(t)}-\XXi_0$.
To linear order in $\delta\XXi(t)$, these are governed by the linear equations $\Dot{\delta \XXi} = C(t)\, \delta \XXi$, where the matrix  $C_{ij}(t) \equiv \partial_j G_i (\XXi_0)$ depends periodically on time with a period $2\pi/\wS$. We remark that since $\w_S$ is a constant of motion $\delta D_{11}^1(t)=\delta D_{11}^1(0)\exp(-\im\wS t)$, which corresponds to a trivial stable evolution. Hence we redefine $\delta\XXi(t)$ as a 12 component vector by removing its $\delta D_{11}^1$  component. Physically these are the fluctuations of the 12 parameters describing the degrees of freedom of a nanomagnet with constant rotational motion about the anisotropy axis.
The time dependence of $C(t)$ can be removed with the following change of variables: $\delta J_\uparrow^r= (\delta J_\downarrow^r)^\dag \equiv \delta {J_\uparrow} \exp[-\im (\varphi -\wS t)]$ and $\delta S_\uparrow^r=(\delta S_\downarrow^r)^\dag \equiv \delta {S_\uparrow} \exp[-\im (\varphi -\wS t)]$.
The linear system reduces then to $\Dot{\delta \XXi^r}= A\, \delta\XXi^r$, where the matrix $A$ is time independent and $\delta \XXi^r$ is obtained replacing the old variables with the new ones, $\delta S_{\uparrow\downarrow}^r$ and $\delta J_{\uparrow\downarrow}^r$.
In the absence of dissipation, linear stability corresponds to the eigenvalues of $A$ being all purely imaginary \cite{meiss2007differential}.

The $12\times12$ complex matrix $A$ can be block-diagonalized as $A_Z\oplus A_T\oplus A_T^*$, where $A_Z$ is a $2\times 2$ matrix defined as
\be\label{eq:Lin_z}
		\pa{t}\begin{pmatrix}
			\delta z\\
			\delta p_z
		\end{pmatrix}
		= A_Z
		\begin{pmatrix}
			\delta z\\
			\delta p_z
		\end{pmatrix}
		\equiv
		\begin{pmatrix}
		0 & 1/M\\
		-M\wZ^2 & 0
		\end{pmatrix}
		\begin{pmatrix}
			\delta z\\
			\delta p_z
		\end{pmatrix},
\ee
where $\wZ$ is defined in \tabref{TAB:Frequencies}.
 $A_T$ is a $5\times 5$ matrix defined as
\be\label{eq:Q_Lin_Sys}
	\pa{t}
	\begin{pmatrix}
		\delta p_+\\
		\delta \rho_+\\
		{\delta J_\uparrow^r}\\
		{\delta S_\uparrow^r}\\
		\delta D_{-10}^{1}
	\end{pmatrix}
	=
	A_T
		\begin{pmatrix}
		\delta p_+\\
		\delta \rho_+\\
		{\delta J_\uparrow^r}\\
		{\delta S_\uparrow^r}\\
		\delta D_{-10}^{1}
	\end{pmatrix},
\ee
where
\begin{widetext}
\be\label{eq:Lin_Matrix}
	A_T \equiv	\im
	\begin{pmatrix}
		0 & -\im \hbar\wL B'' S/(2B_0) &  0 & \im \hbar \gr B' & \im \hbar \gr \sqrt{2} S B'\\
		-\im/M & 0 & 0 & 0 & 0 \\
		0 & \wL S B'/B_0 & \wR +\wS & \wR +\wS - \wL & -\wL\sqrt{2}S \\
		0 & -\wL S B'/B_0 & -\wR & \wL- \wR -2\wD & \wL\sqrt{2}S \\
		0 & 0 & \wR /(\sqrt{2}S) & \wR /(\sqrt{2}S) & 0 \\ 
	\end{pmatrix},
\ee
\end{widetext}
with $ \delta \rho_\pm \equiv \delta x \pm \im \delta y$ and $\delta p_\pm \equiv \delta p_x \pm \im \delta p_y$. The relevant frequencies $\wL,\wR,\wT,\wD$ are defined in \tabref{TAB:Frequencies}.
\begin{table}[t]
\caption{Definition of the relevant frequencies of the system as appearing in \eqnref{eq:Lin_z} and \eqnref{eq:Lin_Matrix}.}\label{TAB:Frequencies}
\begin{ruledtabular}
\begin{tabular}{ll}
Symbol & Definition\\
\hline
$\wR$ & $\hbar S/I = 5\hbar \rho_\mu/(2\mu_B\rho_MR^2)$\\
$\wD$ & $\hbar D S = k_a \mu_B/(\hbar \rho_\mu)$\\
$\wL$ & $\gr B_0$\\
$\wZ$ & $\sqrt{\hbar\gr B'' S/M}$\\
$\wT$ & $\sqrt{\hbar \gr S\pare{B'^2 - B_0B''/2}/{M B_0}}$\\
$\wS$ & $ - \hbar \avg{\Lop_3}/I$\\
\end{tabular}
\end{ruledtabular}
\end{table}
The eigenvalues of $A_Z$, given by the roots of  $P_Z(\lambda) \equiv \lambda^2 + \wZ^2$, are purely imaginary for $B''>0$. This leads to stable harmonic oscillations of the center-of-mass motion along the $\uez$ direction with frequency $\wZ$. 
$A_T$ accounts for the fluctuations of the remaining degrees of freedom and its 
eigenvalues are given by the roots of the fifth order polynomial
\be\label{eq:Poly_NM}
	P_T(\lambda) = a_0 + a_1 \lambda + a_2\lambda^2 + a_3 \lambda^3 + a_4 \lambda^4 + a_5\lambda^5,
\ee
whose coefficients are given by
\be\label{eq:Poly_Gen_Coeff}
\begin{split}
	a_0 \equiv &\, -2\wD\wR \wL \wT^2, \\
	a_1 \equiv &\, \im\spare{ \wD\wZ^2(\wS + \wR) + \wS \wL \wT^2},\\
	a_2 \equiv &\, -2\wD \wR \wL - \frac{1}{2}\pare{2\wD - \wS}\wZ^2 - \wL\wT^2,\\
	a_3 \equiv & \,\im\spare{-2\wD \pare{\wS + \wR} + \wS \wL + \frac{1}{2}\wZ^2},\\
	a_4 \equiv &\, 2\wD - \wS - \wL,\\
	a_5 \equiv &\, - \im.
\end{split}
\ee
This is one of the main results of this article since the roots of $P_Z(\lambda)$ and $P_T(\lambda)$ allow to discern between stable and unstable levitation as a function of the physical parameters of the system via \tabref{TAB:PhysParam} and \tabref{TAB:Frequencies}.
In particular, stable levitation corresponds to the roots of $P_Z(\lambda)$ and $P_T(\lambda)$ being purely imaginary\footnote{One could define $A \equiv \im \tilde A$ such that the characteristic polynomial of $\tilde A$ has real coefficients. Stability would require, in this case, real roots.}.

Let us remark that at the transition between stability and instability the discriminants of $P_Z(\lambda)$ and $P_T(\lambda)$, defined as $\Delta_Z$ and $\Delta_T$ respectively, are zero. This happens whenever two distinct eigenvalues become degenerate (Krein's collision)~\cite{meiss2007differential}.
The eigenvalues of the matrix associated to a linear system of differential equations describing conservative Hamiltonian dynamics, as the matrix $A_Z\oplus A_T \oplus A_T^*$ in our case, are always either complex quadruplets, $\lambda=\{a+\im b,a-\im b,-a+\im b,-a-\im b\}$, real pairs $\lambda=\{a,-a\}$, imaginary pairs $\lambda =\{ \im b, -\im b\}$, or pairs of zero eigenvalues $\lambda=\{0,0\}$, where $a,b\in\mathbb{R}$. Therefore, the transition from stability to instability, namely from all imaginary eigenvalues to have at least a complex quadruplet or a real pair, happens at a Krein's collision. Note that this is a necessary but not sufficient condition since the colliding eigenvalues could still remain on the imaginary axis \cite{meiss2007differential}.

The polynomials $P_Z(\lambda)$ and $P_T(\lambda)$ have also been  obtained via the classical description of the nanomagnet discussed in \secref{sec:ClassDesc}.
The procedure is very similar to the one presented above, but care must be taken when linearizing the system around the solution represented in \figref{FIG:Frequencies}.a since it corresponds to a degeneracy point of the Euler angular coordinates in the ZYZ convention. That is, for $\eulb=0$ it is not possible to distinguish between rotation of the angle $\eula$ and $\eulc$. This is the so-called Gimbal lock problem which can be circumvented either by using an alternative definition of the Euler angles, which moves the degeneracy point elsewhere, or by changing the parametrization of the Ioffe-Pritchard field \eqnref{eq:Ioffe_Pritchard}, namely by aligning the bias along the $\uex$- or $\uey$-axis.
The Gimbal lock problem is avoided in the quantum description in the frame $\Ob$ by the use of the D-matrices.

\section{Linear stability diagrams}\label{sec:NM_Stab}

Using the criterion derived in \secref{sec:LinStab}, let us now analyze the linear stability of the nanomagnet at the equilibrium point illustrated in \figref{FIG:Frequencies}.a (nanomagnet at the center of the trap anti-aligned to the local magnetic field) as a function of the physical parameters given in \tabref{TAB:PhysParam} and the rotation frequency $\wS$.

As shown below the stability of the system depends very much on the size of the magnet, parametrized by $\wR$, the local magnetic field strength, parametrized by $\wL$, and the magnetic rigidity given by the magnetic anisotropy energy, parametrized by $\wD$. 
In particular, we distinguish the following three regimes: (i) the small hard magnet (sHM) regime, $\wR,\wD\gg \wL$, (ii) the soft magnet regime (SM), $\wD\ll\wL$, and (iii) the large hard magnet (lHM) regime, $\wD\gg\wL\gg\wR$.

\begin{figure*}
	\includegraphics[width= 2\columnwidth]{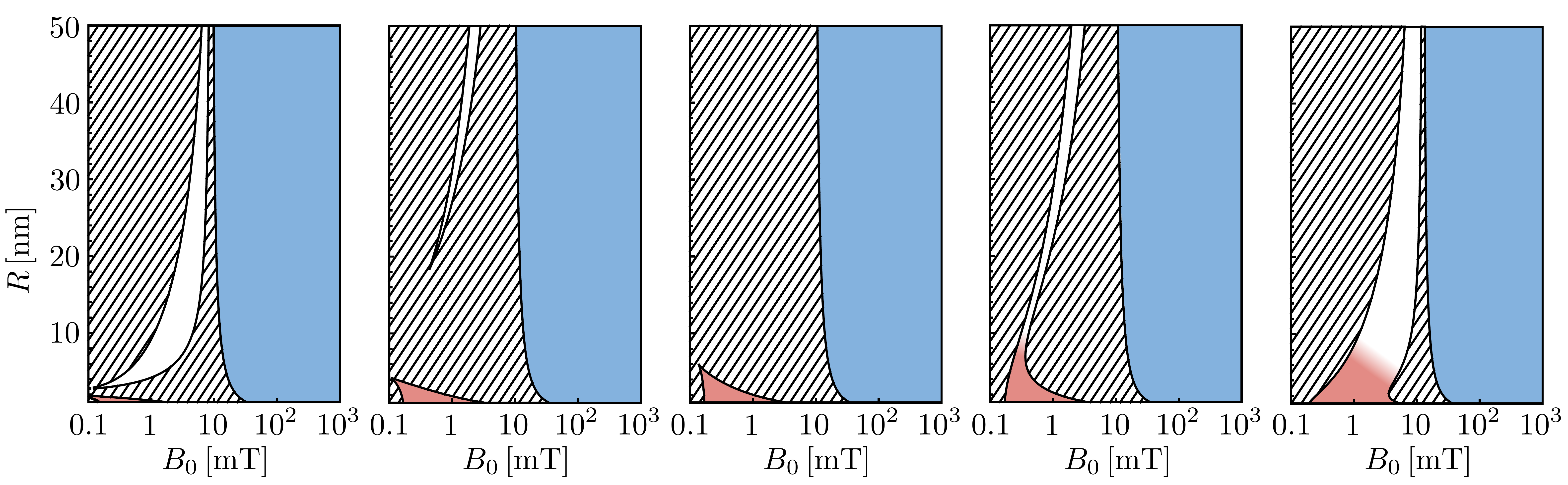}
	\caption{From left to right stability diagrams for $\wS/\w_0 = -0.2,\,-0.02,\, 0,\, 0.02, \, 0.2$, where $\w_0/(2 \pi) \approx 5 \times 10^{8}~\text{Hz}$ is $\wR$ for a nanomagnet of radius $R=1\,\text{nm}$. Other physical parameters are taken from the caption of \tabref{TAB:PhysParam}. 
	Stable phases are illustrated in red (EdH phase), blue (A phase), and white (L phase). $\wS>0$ $(\wS<0)$ corresponds to  clockwise rotation (counter clockwise rotation).}
	\label{FIG:PhaseDiag_NM} 
\end{figure*}

We present the results in a two-dimensional phase diagram with the $x$-axis given by the bias field $B_0$ and the $y$-axis given by the radius of the nanomagnet $R$. Results are shown in \figref{FIG:PhaseDiag_NM}. Note that in the sHM (lower left corner) and SM (right part) regimes  two stable phases are present for a non-rotating nanomagnet ($\wS=0$) (central panel). In the lHM regime (upper left corner) on the other hand, stable levitation is possible only for a mechanically rotating nanomagnet ($\wS\neq0$). As argued below these three stable phases have a different physical origin and represent three different loopholes in the Earnshaw theorem, the Einstein-de Haas (EdH) loophole, the atom (A) loophole, and Levitron (L) loophole.

\subsection{Einstein-de Haas phase}

In the sHM regime, where $\wD\gg\wL$, the magnetic moment can be considered, to a good approximation, fixed along the direction of the magnetic anisotropy.
Due to the small dimension of the nanomagnet the spin angular momentum plays a significant role in the dynamics of the system, namely $\wR\gg\wL$ (see \figref{FIG:Frequencies}.b).
$2\wR$ is indeed the frequency at which the nanomagnet would rotate if the magnetic moment flipped direction, in accordance with the Einstein-de Haas effect \citep{EinsteindeHaas}.
Such effect thus plays a relevant role in the dynamics of the system in the sHM regime due to the small moment-of-inertia-to-magnetic-moment ratio.
In particular, a strong EdH effect, \ie a large $\wR$ compared to the other frequencies of the system, has the effect of locking the quantum spin along one of the anisotropy directions due to energy conservation \citep{Chud_EdH1,Chud_RotStatesNM,O'Keeffe_PRB}. In the absence of rotation, the spin-rotation interplay described by the Einstein--de Haas effect thus stabilizes the non-rotating magnet by keeping the macrospin aligned along the anisotropy direction.

\begin{figure}
	\includegraphics[width= 0.9\columnwidth]{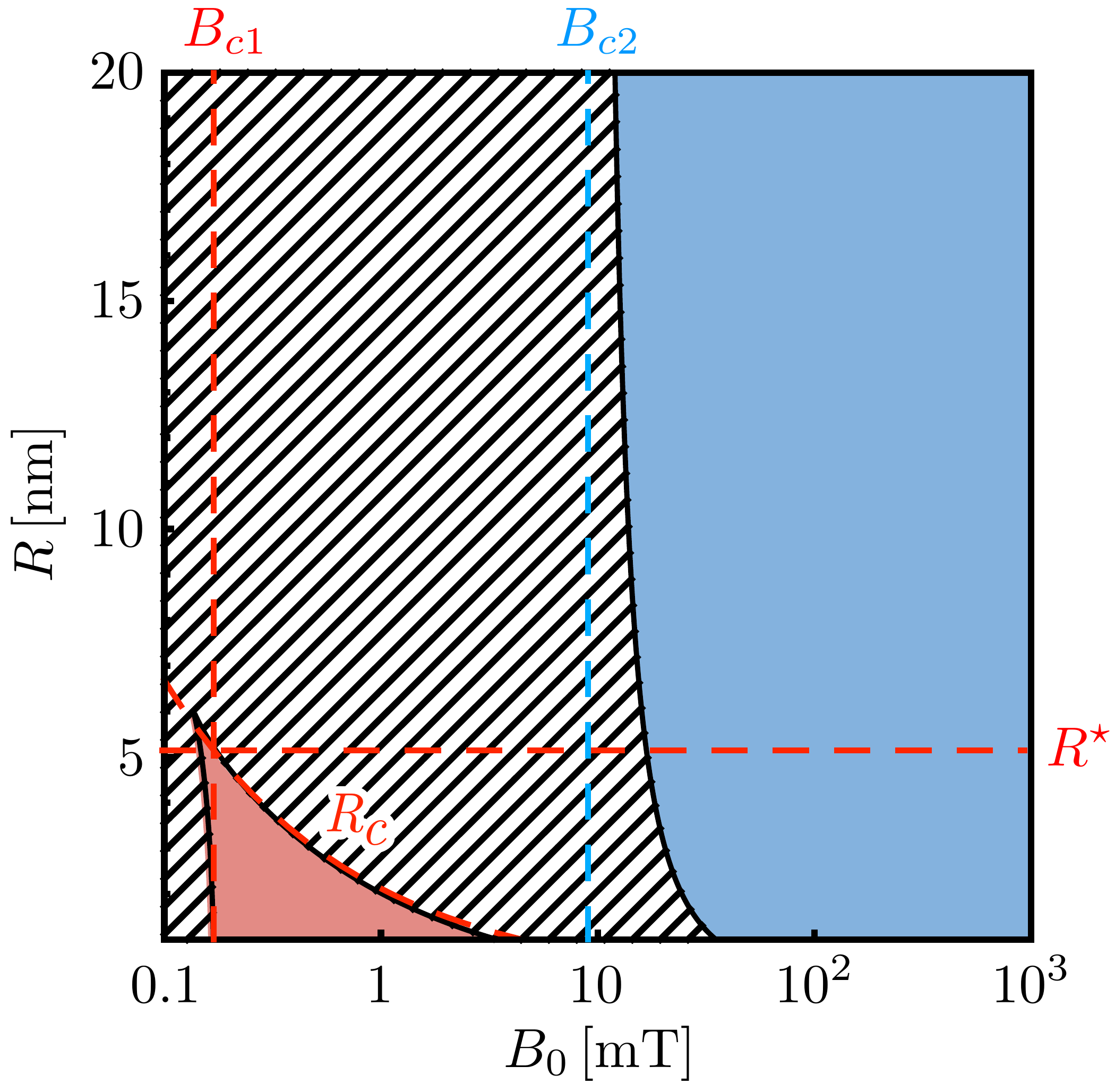}
	\caption{Stability diagram for a non-rotating nanomagnet ($\wS=0$). Other physical parameters are given in the caption of \tabref{TAB:PhysParam}. The approximated borders of the red EdH phase (blue A phase) are illustrated with red (blue) dashed lines.}
	\label{FIG:Analytical_Borders} 
\end{figure}

\begin{figure*}
	\includegraphics[width= 2\columnwidth]{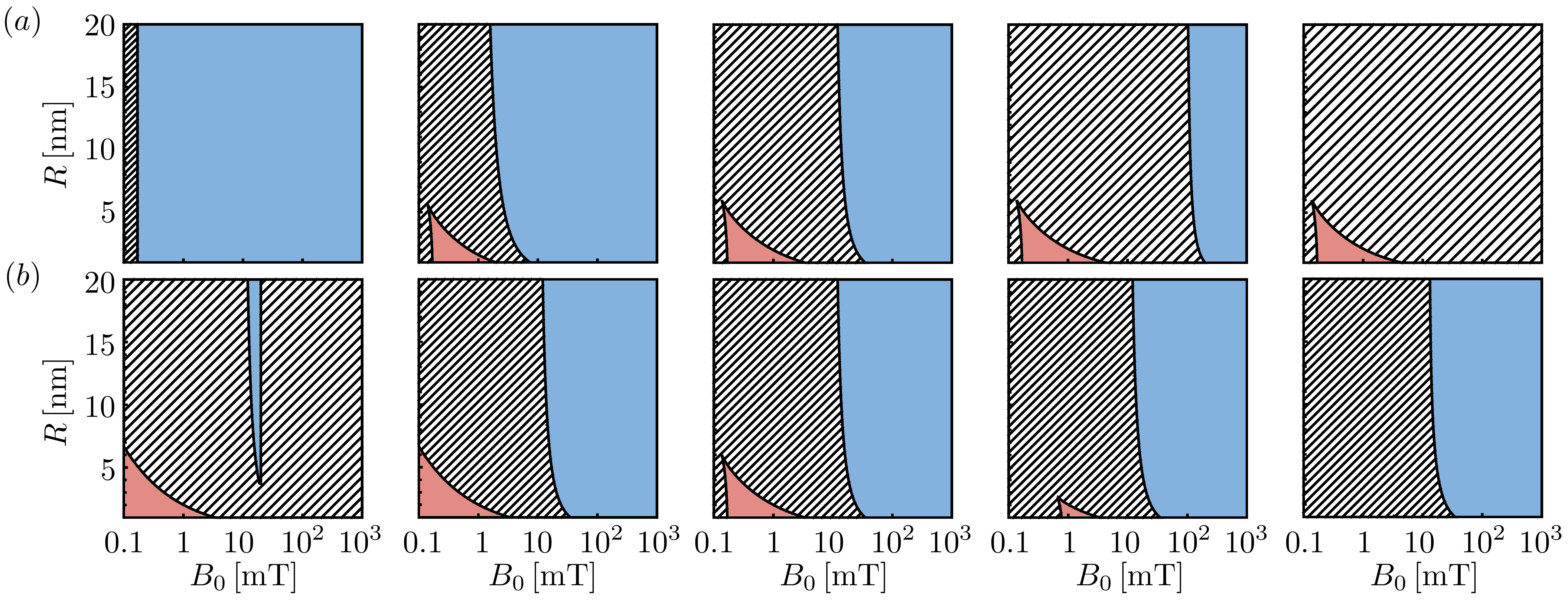}
	\caption{From left to right stability diagrams for a non-rotating nanomagnet ($\w_S=0$) for (a) $k_a [\text{J}/\text{m}^3]=0,10^3,10^4,10^5,\infty$ and for (b)  $B'\,[\text{T}/\text{m}]=10^2, 10^3, 10^4, 10^5, 10^6$. Other physical parameters are given in the caption of \tabref{TAB:PhysParam}. Stable phases are illustrated in red (EdH phase) and blue (A phase).}
	\label{FIG:PhaseDiag_QualitativeBehavior} 
\end{figure*}

The borders of the EdH stable phase in the non-rotating case can be analytically approximated as follows, see \figref{FIG:Analytical_Borders}.
The upper border can be approximated by keeping terms in $\Delta_T=0$ of zero order in $\wZ/\wD\ll1$ and up to leading order in $\w/\wD\ll1$ (for  $\w=\wT,\wL,\wR$).  This is justified in the sHM regime, see \figref{FIG:Frequencies}.b.
This leads to the simple expression $\wR=4\wL$, which using \tabref{TAB:Frequencies}, reads
\be\label{eq:R_c}
	R_c \equiv \sqrt{\frac{5\rho_\mu}{8\gr^2 B_0 \rho_M}}.
\ee
Given $B_0$, \eqnref{eq:R_c} approximates the maximum radius to allow stable levitation. Such an approximated expression is in good agreement with the exact upper border, see \figref{FIG:Analytical_Borders}.
The left border can be approximated by keeping terms in $\Delta_T=0$ of zero order in $\wZ/\wD\ll1$ and of highest order in $\wR/\wD\gg1$, which is justified in the sHM regime for $R \rightarrow 0$, see \figref{FIG:Frequencies}.b. This leads to $\wL = 3\sqrt{3}\wT/2$, which using \tabref{TAB:Frequencies}, reads
\be\label{eq:B0_star}
     B_{c1} \equiv 3\pare{\frac{\rho_\mu B'^2}{4 \gr^2 \rho_M}}^{\frac{1}{3}},
\ee
where we neglected the contribution $B''$ in $\wT$, since $B''B_0/B'^2 \ll 1$. 
This approximates the minimum $B_0$ for stable levitation in the EdH phase. As shown in \figref{FIG:Analytical_Borders}, this gives a good estimation of the left border. Plugging \eqnref{eq:B0_star} into \eqnref{eq:R_c} one obtains an approximated expression for the radius $R^\star$ of the largest nanomagnet that can be stably levitated in the non-rotating EdH phase. Note that these expressions explain the dependence of the EdH phase on the field gradient $B'$ and the uniaxial anisotropy strength $k_a$ shown in \figref{FIG:PhaseDiag_QualitativeBehavior}.

In particular, note that the EdH phase is nearly independent of $k_a$ provided the condition $\w_D \gg \w_L$ holds. Therefore, one can describe this regime with a simplified model assuming $k_a \rightarrow \infty$ (perfect hard magnet), which corresponds to the magnetic moment frozen along $\bold{e}_3$ (rightest panel in \figref{FIG:PhaseDiag_QualitativeBehavior}.a). 
In this limit, the Hamiltonian of the system reads
\be\label{eq:Ham_HM_General}
	\Hop_\text{sHM} =  \frac{\hat{\pp}^2}{2M} + \frac{\hbar^2}{2I}\pare{\JJop^2 + 2S \Jop_3} +\hbar\gamma_0 S B_3(\hat \rr,\Eang).
\ee
\eqnref{eq:Ham_HM_General} is obtained from \eqnref{eq:Ham_Body_General} projecting the spin degrees of freedom on the eigenstate $\ket{S}$ of $\Sop_3$, where $S$ is the largest value for the spin projection along $\bold{e}_3$. In the classical description, this limit corresponds to the Lagrangian
\be\label{eq:LL_HM_Lagrangian}
\begin{split}
	\mathcal{L}_\text{sHM} =&  \frac{I}{2}\spare{\deula^2+\deulb^2+\pare{\deulc+\wR}^2+2\pare{\deulc+\wR}\deula\cos\beta}\\
	&+\frac{M}{2}\pare{\Dot{x}^2+\Dot{y}^2+\Dot{z}^2} +\mu\big[B_x(\rr)\cos\eula\sin\eulb \\
	&+B_y(\rr)\sin\eula\sin\eulb +B_z(\rr)\cos\eulb \big],
\end{split}	
\ee
where we set $\theta = \beta$ and $\phi=\eula$. \eqnref{eq:LL_HM_Lagrangian} shows that $\wR$ appears as a shift of the rotational frequency of the nanomagnet about $\bold{e}_3$. This shift, which must not be interpreted as an actual mechanical rotation, represents the contribution of the macrospin to the total angular momentum of the system. This effect can also be seen in the characteristic polynomial, see \tabref{TAB:Poly_Coeff}.
The linear stability analysis using \eqnref{eq:Ham_HM_General} or \eqnref{eq:LL_HM_Lagrangian} leads to $P_Z(\lambda)$ (as in the general case) but to a simplified $P_T(\lambda)$ given by
\be\label{eq:Poly_HM}
	P^\text{sHM}_T(\lambda) = a_0^\text{sHM} + a_1^\text{sHM} \lambda + a_2^\text{sHM} \lambda^2 + a_3^\text{sHM} \lambda^3 + a_4^\text{sHM} \lambda^4,
\ee
whose coefficients are given in \tabref{TAB:Poly_Coeff}. This leads to the stability diagram shown in the rightest panel of 
\figref{FIG:PhaseDiag_QualitativeBehavior}.a. Note that $P^\text{sHM}_T(\lambda)$ is of fourth order since the magnetization is frozen along $\bold{e}_3$ and hence there are only $10$ independent parameters.

 \begin{table}
 \caption{Coefficients of the stability polynomial $P_T(\lambda)$ in the sHM, lHM and SM regime.}\label{TAB:Poly_Coeff}
 \begin{ruledtabular}
 \begin{tabular}{l|lll}
 	& sHM ($a_i^\text{sHM}$) & SM ($a_i^\text{SM}$) & lHM ($a_i^\text{lHM}$)  \\
 	\hline
 	$a_0$ & $-\wR\wL\wT^2$ &  $-\wL\wT^2$  & $-\wR\wL\wT^2$\\
 	$a_1$ & $\im\wZ^2(\wS+\wR)/2$ &  $\im\wZ^2/2$ & $\im\wZ^2\wS/2$\\
 	$a_2$ & $-\wR\wL-\wZ^2/2$ & $-\wL $ & $-\wR\wL-\wZ^2/2$\\
 	$a_3$ & $-\im(\wS+\wR)$ & $-\im$ &  $-\im\wS$\\
 	$a_4$ & $1$ & $0$ & $1$  
 \end{tabular}
 \end{ruledtabular}
 \end{table}

\subsection{Atom phase}

In the SM regime, where $\wL\gg\wD$ (see \figref{FIG:Frequencies}.b), the coupling between the magnetization and the anisotropy is negligible. In this regime, the magnetic moment undergoes a free Larmor precession about the local magnetic field. This stabilizes the system in full analogy to magnetic trapping of neutral atoms \cite{Sukumar1997,Sukumar2006}.

The borders of the A phase are approximately independent on the rotational state of the nanomagnet $\wS$, as shown in \figref{FIG:PhaseDiag_NM}. Therefore, considering the case of a non-rotating nanomagnet, they can be analytically approximated as follows, see \figref{FIG:Analytical_Borders}.
The left border at low magnetic fields can be approximated by keeping only terms in $\Delta_T=0$ up to zero order in $\wZ/\wL \ll 1$ and up to leading order in $\wR/\wL \ll 1$ and $\wT/\wL \ll 1$, which is well justified in the SM regime at $R \rightarrow \infty$, see \figref{FIG:Frequencies}.b. This leads to the condition $\wL=2\wD$, which using \tabref{TAB:Frequencies} reads
\be\label{eq:B0_Star2}
    B_{c2} \equiv \frac{2 k_a}{\rho_\mu}.
\ee
$B_{c2}$ approximates the lowest field bias for which stable levitation is possible in the A phase, see \figref{FIG:Analytical_Borders}. 
The A phase extends up to the field bias $B_{c3} = 2B'^2/B''$, above which $\wT$ becomes imaginary. This is shown in the leftmost diagram in \figref{FIG:PhaseDiag_QualitativeBehavior}.b, while in the remaining panels it falls out of the $B_0$ interval shown. Note that there is no upper limit in $R$ for the A phase. However recall that our model assumes a single magnetic domain nanomagnet, which most materials can only sustain for sizes up to hundred nanometers~\cite{PrincipleNanomagnetism}. 
Note that the dependence of $B_{c2}$ and $B_{c3}$ on the field gradient $B'$ and the uniaxial anisotropy strength $k_a$ explains the qualitative behaviour of the A phase in \figref{FIG:PhaseDiag_QualitativeBehavior}.

In the limit of a vanishing magnetic anisotropy, $k_a=0$, the Hamiltonian of the nanomagnet reads $\Hop_\text{SM} = \Hop_\text{AT} + \hbar^2\LLop^2/2I$, where $\Hop_\text{AT} = \ppop^2/2M - \hbar\gr \FFop\cdot\BB(\rrop)$ represents the Hamiltonian describing a single magnetic atom of mass $M$ and spin $F$ in the external field $\BB(\rr)$ \cite{Sukumar1997,Sukumar2006}.
In the same limit, the system is described classically by the Lagrangian $\mathcal{L}_\text{SM}$ obtained from $\mathcal{L}$ by setting $V_a=0$, thus decoupling rotation and magnetization dynamics.
In this limit, the linear stability analysis applied to $\Hop_\text{SM}$ or to $\mathcal{L}_\text{SM}$ leads to $P_Z(\lambda)$ (as in the general case) and to 
\be\label{eq:Poly_SM}
	P^\text{SM}_T(\lambda) = a_0^\text{SM} + a_1^\text{SM}\lambda + a_2^\text{SM}\lambda^2 + a_3^\text{SM} \lambda^3,
\ee
whose coefficients given in \tabref{TAB:Poly_Coeff} are, as expected, independent on $\wS$ and $\wR$, namely on the rotational state of the nanomagnet.
This leads to the stability diagram shown in the leftmost panel in \figref{FIG:PhaseDiag_QualitativeBehavior}.a, whose left border coincides with \eqnref{eq:B0_star}.
Note that $P_T^\text{SM}(\lambda)$ is only a third order polynomial because the rotational dynamics do not affect the stability of the system. The only relevant degrees of freedom for the stability are thus the magnetic moment and the center-of-mass motion (8 independent parameters).

\subsection{Levitron phase}

In the lHM regime, the magnetic moment can be considered to be fixed along the easy axis ($\wD \gg \wL$) and the contribution of the spin to the total angular momentum can be neglected due to the large dimension of the nanomagnet ($\wL \gg \wR$), see \figref{FIG:Frequencies}.b.  In this respect, the nanomagnet behaves in good approximation like a classical Levitron.
The dynamics in this regime can be approximately described by the Hamiltonian
\be\label{eq:Ham_Levitron}
	\Hop_\text{lHM} = \frac{\ppop^2}{2M} + \frac{\hbar^2}{2I} \LLop^2 - \mu \bold{e}_3(\Eang)\cdot \BB(\rrop),
\ee
which is obtained from $\Hop$ by taking the limit $k_a \rightarrow \infty$ (magnetization frozen along the anisotropy axis) and by using $\hat{\mmu}=\mu\bold{e}_3(\Eang)$. The latter treats the magnetization classicaly, namely $\mu$ is a scalar quantity  instead of a quantum spin operator.
The classical description is given in this limit by the Lagrangian
\be\label{eq:LL_lHM_Lagrangian}
\begin{split}
	\mathcal{L}_\text{lHM} = & \frac{I}{2}\pare{\deula^2+\deulb^2 + \deulc^2 + 2\deula\deulc \cos\eulb}\\
	&+\frac{M}{2}\pare{\Dot x ^2 + \Dot y^2 + \Dot z^2} + \mmu\cdot \BB(\rr),
\end{split}
\ee
where $\mmu=\mu(\cos\eula\sin\eulb,\sin\eula\sin\eulb,\cos\eulb)$ for a magnetic moment frozen along the anisotropy axis.

The linear stability analysis applied to this limit leads to the polynomials $P_Z(\lambda)$ (as in the general case) and $P^\text{lHM}_T(\lambda)=a_0^\text{lHM} + a_1^\text{lHM} \lambda + a_2^\text{lHM}\lambda^2 + a_3^\text{lHM} \lambda^3 + a_4^\text{lHM} \lambda^4$, where its coefficients are defined in \tabref{TAB:Poly_Coeff}.
The linear stability diagram derived from $P_Z(\lambda)$ and $P^\text{lHM}_T(\lambda)$ corresponds to the L phase of the lHM regime in \figref{FIG:PhaseDiag_NM}, thus showing that stable levitation in this regime requires mechanical rotation. 
Furthermore, in this limit the stability region is symmetric with respect to clock- or counterclockwise rotation, as in the classical Levitron \cite{Gov1999214,Dullin,Berry_Levitron}.

To conclude this section, let us compare the description of the magnetic moment in the approximated models of the lHM and the sHM regimes. 
The lHM and sHM both describe a nanomagnet with a large magnetic rigidity whose magnetic moment can be approximated to be frozen along the easy axis $\bold{e}_3$. 
In the lHM regime, due to the negligible role played by the macrospin angular momentum ($\wR\ll\wL$), the magnetic moment is modeled as $\hat{\mmu}=\mu\bold{e}_3(\Eang)$, where $\mu$ is a classical scalar quantity.
In the sHM regime, on the other hand, the role of the spin angular momentum is crucial ($\wR\gg\wL$) , and the quantum origin of the nanomagnet's magnetic moment has to be taken into account. The magnetic moment is thus given by $\hat{\mmu}=\hbar \gr [\FFop\cdot\bold{e}_3(\Eang)]\bold{e}_3(\Eang)$.
This crucial difference is manifested in the coefficients of the characteristic polynomial, see \tabref{TAB:Poly_Coeff}. In the sHM regime the rotational frequency $\wS$ is shifted by $\wR$, thus retaining the contribution of the spin angular momentum $\FFop$ to the total angular momentum of the system. In essence, the quantum spin origin of the magnetization plays the same role as mechanical rotation, a manifestation of the Einstein-de Haas effect.

\section{Conclusions}\label{sec:Conclusions}

In conclusion, we discussed the linear stability of a single magnetic domain nanosphere in a static external Ioffe-Pritchard magnetic field
at the equilibrium point illustrated in \figref{FIG:Frequencies}.a. This corresponds to a nanomagnet at the center of the field, with the magnetic moment parallel to the anisotropy axis, anti-aligned to the magnetic field, and mechanically spinning with a frequency $\w_S$.
We derived a stability criterion given by the roots of both a second order polynomial $P_Z(\lambda)$ and of a fifth order polynomial $P_T(\lambda)$. Eigenvalues with zero (non-zero) real component  correspond to stability (instability). This stability criterion is derived both with a quantum description and a (phenomenological) classical description of the nanomagnet. 
Apart from the known gyroscopic-stabilized levitation (Levitron L phase), we found two additional stable phases, arising from the quantum mechanical origin of the magnetization, $\hat{\mmu}=\hbar \gr \FFop$, which surprisingly (according to Earnshaw's theorem) allow to stably levitate a non-rotating magnet.
The atom A phase appears at a high magnetic bias field ($\wL\gg \wD$), where despite the magnetocrystalline anisotropy the magnetic moment freely precesses along the local direction of the magnetic field. The stability mechanism is thus fully analogous to the magnetic trapping of neutral atoms \cite{Sukumar1997,Sukumar2006}.
The Einstein-de Haas EdH phase arises at a low magnetic bias field ($\wL\ll\wD$), where the uniaxial magnetic anisotropy interaction dominates the magnetization's dynamics. The magnetic moment is thus frozen along the easy axis and can be modeled as $\hat{\mmu}=-\hbar\gr[\FFop\cdot\bold{e}_3(\Eang)]\bold{e}_3(\Eang)$.
In this case the quantum spin origin of $\mmu$ is crucial to stabilize the levitation of a small nanomagnet through the Einstein-de Haas effect.
As the size of the nanomagnet increases, the contribution of the spin angular momentum becomes negligible due to the increasing moment-of-inertia-to-magnetic-moment ratio and the classical Levitron behavior is recovered.

To derive these results, we assumed a: (i) single magnetic domain, (ii) macrospin approximation, (iii) rigid body, (iv) sphere, (v) uniaxial anisotropy, (vi) Ioffe-Pritchard magnetic field, (vii) point-dipole approximation, (viii) that gravity can be neglected, namely $M g / (\mu B'') \ll (B_0/B'')^{1/2}, (B'/B'')$ , and (ix) dissipation-free dynamics for the system. 
While not addressed in this article, it would be very interesting to relax some of these assumptions and study their impact on the stability diagrams.
For instance, levitating a multi-domain magnet could allow  to study the effects of the interactions between different domains on the stability of the system. 
It would be particularly interesting to explore if the A phase persists for a macroscopic multi-domain magnet at sufficiently high magnetic fields. In this scenario and depending on the size of the magnet, not only assumption (i) and (ii), but also (iii), (v), (vii), (viii) and (ix) should be carefully revisited.
One could use the exquisite isolation from the environment obtained in levitation in high vacuum to study in-domain spin dynamics beyond the macrospin approximation.
Generalization to different shapes and magnetocrystalline anisotropies would allow to investigate the shape-dependence of the stable phases, as done for the Levitron~\cite{Gov1999214}. In particular, one could explore the presence of multi-stability with other magnetocrystalline anisotropies that contain more than a single easy axis.
Levitation in different magnetic field configurations, such as quadrupole fields, might be used to further study the role of $B_0$ (crucial for the levitation of neutral magnetic atoms \cite{Sukumar1997,Sukumar2006}) in the levitation of a nanomagnet. In particular, to discern whether stable levitation can occur in a position where the local magnetic field is zero.
The effect of noise and dissipation on the stability of the system might not only enrich the stability diagram, but also play a crucial role in any experiments aiming at controlling the dynamics of a levitated nanomagnet.
We remark that linear stability is a necessary but not sufficient condition for the stability of the system at long time scales. A thorough analysis of the stability of a nanomagnet in a magnetic field under realistic conditions might demand to consider non-linear dynamics.

To conclude, we remark that one could consider to cool the fluctuations of the system in the stable phases to the quantum regime. 
The fluctuations of the degrees of freedom of the system could then be described as coupled quantum harmonic oscillators using the bosonization tools given in \cite{ORI_Cos_PRB}. This procedure leads to a quadratic bosonic Hamiltonian describing the dynamics around the equilibrium point. The linear equations for the bosonic modes yield the same characteristic polynomials $P_Z(\lambda)$ and $P_T(\lambda)$ derived in this article within the classical and semiclassical approach. Moreover, the bosonization approach allows to study the quantum properties (entanglement and squeezing) of the relevant eigenstates of the quadratic bosonic Hamiltonian~\cite{Short_Paper}, and exploit the rich physics of magnetically levitated nanomagnets in the quantum regime.

We thank K. Kustura for carefully reading the manuscript.
This work is supported by the European Research Council (ERC-2013-StG 335489 QSuperMag) and the Austrian Federal Ministry of Science, Research, and Economy (BMWFW).

\end{document}